\documentclass{PoS}
\newcommand{\beqn}{\begin{eqnarray}}
\newcommand{\eeqn}{\end{eqnarray}}

\title{Study of thermal monopoles in lattice QCD}

\ShortTitle{Study of thermal monopoles in lattice QCD}

\author{\speaker{Vitaly Bornyakov}%
\\
        Institute for High Energy Physics, 142280, Protvino, Russia\\
        E-mail: \email{vitaly.bornyakov@ihep.ru}}

 \author{A.G. Kononenko\\
        Moscow State University, Physics Department, Moscow, Russia \\
        E-mail: \email{agkono@gmail.com}}

\author{V.K. Mitrjushkin\\
        Joint Institute for Nuclear Research, 141980, Dubna, Russia and \\
        Institute of Theoretical and Experimental Physics, 117259 Moscow, Russia\\
        E-mail: \email{vmitr@theor.jinr.ru}}

\abstract{The properties of the thermal Abelian color-magnetic
monopoles in the maximally Abelian gauge are studied in the
vicinity of the confinement-deconfinement phase transition in the
lattice $SU(3)$ gluodynamics and lattice QCD. We compute the
density  and interaction parameters of the thermal
 monopoles. We find that the density of the thermal
monopoles $\rho(T)$ jumps up near the transition temperature $T_c$. 
Additionally we
present new results on the percolation transition in $SU(3)$ gluodynamics 
which is known to coincide in
gluodynamics with the confinement-deconfinement phase transition.
 }

\FullConference{31st International Symposium on Lattice Field Theory LATTICE 2013\\
                 July 29 – August 3, 2013\\
                 Mainz, Germany}

\begin{document}

\section{Introduction}

The nonperturbative properties of the nonabelian gauge theories,
e.g., confinement, confinement-deconfinement transition, chiral symmetry
breaking, etc. are closely related to the Abelian monopoles
defined in the maximally Abelian gauge
$(MAG)$~\cite{'tHooft:1981ht,Kronfeld:1987ri}.

It has been recently argued that the MAG is a proper Abelian gauge
to find gauge invariant monopoles since t'Hooft-Polyakov monopoles
can be identified in this gauge by the Abelian flux, but this is
not possible in other Abelian gauges~\cite{Bonati:2010tz}.  

In recent papers~\cite{Liao:2006ry,Chernodub:2006gu} it has been
suggested that color-magnetic monopoles contribution can explain
the strong interactions in the quark-gluon matter which were found
in heavy ion collisions experiments~\cite{Adams:2005dq}.  These
proposals inspired  studies of the properties and possible roles
of the monopoles in the quark-gluon
phase~\cite{Shuryak:2008eq,Ratti:2008jz,D'Alessandro:2007su,Liao:2008jg,Chernodub:2009hc,
D'Alessandro:2010xg,Bornyakov:2011th,Bornyakov:2011eq,Bornyakov:2011dj,Braguta:2012zz}.

In Ref.~\cite{Chernodub:2006gu} it has been shown that thermal
monopoles in Minkowski space are associated with Euclidean
monopole trajectories wrapped around the temperature direction of
the Euclidean volume.  So the density of the monopoles in the
Minkowski space is given by the average of the absolute value of
the monopole wrapping number. First numerical investigations of
the wrapping monopole trajectories were performed in $SU(2)$
Yang-Mills theory at high temperatures in
Refs.~\cite{Bornyakov:1991se} and~\cite{Ejiri:1995gd}.~A more
systematic study of the thermal monopoles was performed in
Ref.~\cite{D'Alessandro:2007su}. It was found
in~\cite{D'Alessandro:2007su} that the density of monopoles is
independent of the lattice spacing, as it should be for a physical
quantity.  The density--density spatial correlation functions were
also computed in~\cite{D'Alessandro:2007su}.  It was shown that
there is a repulsive (attractive) interaction for a
monopole--monopole (monopole--antimonopole) pairs, which at large
distances might be described by a screened Coulomb potential with
a screening length of the order of $0.1$ fm.  In
Ref.~\cite{Liao:2008jg} it was proposed to associate the
respective coupling constant with a magnetic coupling
$\alpha_{m}$.  In the paper~\cite{D'Alessandro:2010xg}
trajectories which wrap more than one time around the time
direction were investigated. It was shown that these trajectories
contribute significantly to a total monopole density at $T$
slightly above $T_c$.  It was also demonstrated that Bose
condensation of thermal monopoles, indicated by vanishing of the
monopole chemical potential, happens at temperature very close to
$T_c$.

The quantitatively precise determination of  such parameters as
monopole density, monopole coupling and others is necessary, in
particular, to verify the conjecture ~\cite{Liao:2006ry} that the
magnetic monopoles are weakly interacting (in comparison with
electrically charged fluctuations) just above transition but
become strongly interacting at high temperatures.

\vspace{2mm} So far lattice studies of the thermal monopoles
were mostly made for $SU(2)$ gluodynamics. First results for 
$SU(3)$ gluodynamics and QCD were presented in our previous
paper  \cite{Bornyakov:2012gqa}
In this paper we present new results of our study of the thermal
monopoles in the $SU(3)$ gluodynamics concentrating on the vicinity of
the confinement-deconfinement phase transition.
To avoid systematic effects due to Gribov
copies we use the gauge fixing procedure as in
Ref.~\cite{Bornyakov:2003vx} with 10 gauge copies.

In paper \cite{Bonati:2013bga} the thermal monopoles were investigated with new
definition of MAG. The results obtained were in qualitative agreement 
with results of Ref.~\cite{Bornyakov:2012gqa}.

\section{Definitions and simulation details }

MAG is determined by the gauge condition \cite{'tHooft:1981ht}
\begin{equation}
\sum_{c \ne 3,8} \left( \partial_\mu \delta_{ac} + \sum_{b=3,8}
f_{abc} A^b_\mu(x) \right) A^c_\mu(x)=0\,,\quad a \ne 3,8
\end{equation}
Solutions of this equation are extrema (over $g$) of the
functional $ F_{\mathrm{MAG}}[A^g] $
\begin{equation}
 F_{\mathrm{MAG}}[A] = \frac{1}{V} \int d^4 x \, \sum_{a \ne 3,8} [A^a_\mu(x)]^2
\end{equation}
Abelian projection means discarding offdiagonal components from the observables 
\begin{equation}
 \sum_a A^a_\mu(x) T^a \to A^3_\mu(x) T^3 + A^8_\mu(x) T^8\\
\end{equation}

On a  lattice MAG gauge fixing functional and Abelian projection
are of the form \cite{Kronfeld:1987ri}
\begin{equation}
F(U) = \frac{1}{V} \sum_{x,\mu,i}   \left( |U_\mu(x)^{ii}|^2 
 \right)\,,\,\,\,\,\,
 U_\mu(x) \to u_\mu(x) \in U(1)^2
 \end{equation}
After Abelian projection one can define magnetic currents: 
\beqn
j_{\mu}^{i} \equiv  \frac{1}{4\pi}\epsilon_{\mu\nu\rho\sigma}
\partial_{\nu} \overline{\Theta}_{\rho\sigma}^{(i)} = -\frac{1}{2}
\epsilon_{\mu\nu\rho\sigma} \partial_{\nu} m_{\rho\sigma}^{(i)}\,,
i=1,2,3 \label{eq:mon:current} 
\eeqn 
were
$\overline{\Theta}_{\rho\sigma}^{(i)}$ is lattice Abelian field
strength. The magnetic currents satisfy the constraint \beqn
\sum_a j_{\mu}^{i}(x) = 0\,, \label{eq:k-constraint} \eeqn on
any link $\{x,\mu\}$ of the dual lattice. Furthermore magnetic
currents are conserved and form closed loops.

Thermal monopoles are defined as clusters of magnetic currents
wrapped in the temperature dimension. Wrapping number for given cluster
$N^i_{wr}$ is equal to:
\begin{equation}
 N^i_{wr} = \frac{1}{3L_t} \sum_{j^i_4(x) \in cluster} j^i_4(x)
\end{equation}
Then respective density is
\begin{equation}
\rho = \frac{\langle~ \sum_{clusters,a}|N^i_{wr}| ~\rangle }{3L_s
^3 a^3}
\label{def_rho}
\end{equation}
One can also define the densities $\rho_k$ of the thermal monopoles wrapped
$k$ times.

$SU(3)$ lattice gluodynamics was simulated with Wilson action. 
Configurations of $N_f=2$ lattice QCD were produced on lattices
$32^3 \cdot 12$ with
non-perturbatively $O(a)$ improved Wilson fermionic action at
 $\beta=5.25$. It had been found   by
DIK collaboration \cite{Bornyakov:2009qh} that at crossover
$T_c \approx 200$ MeV and $m_\pi \approx  400 $MeV.

\section{Results}

\begin{figure}[h]
\begin{center}
\includegraphics[angle=270,scale=0.4]{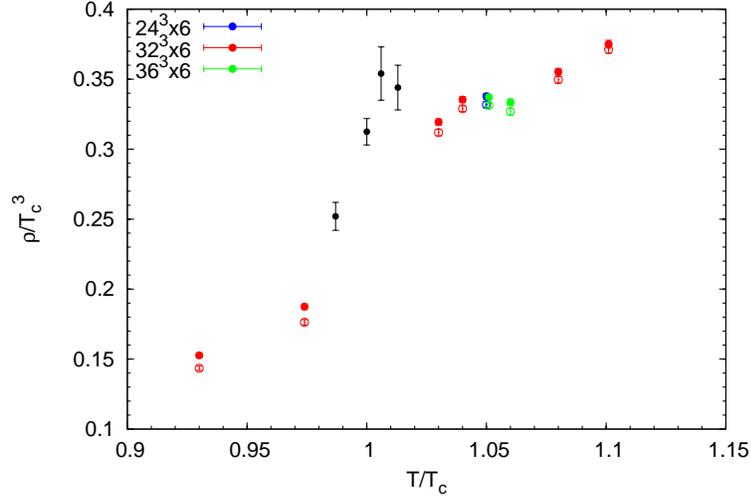}\\
\caption{Full monopole density $\rho(T)$ (full symbols) and 
density $\rho_1(T)$ (empty symbols)  in
SU(3) gluodynamics vs $T/T_c$. Full monopole density for full
QCD (black symbols) is shown for comparison.} \label{fig1}
\end{center}
\end{figure}

\begin{figure}[h]
\begin{center}
\includegraphics[angle=270,scale=0.4]{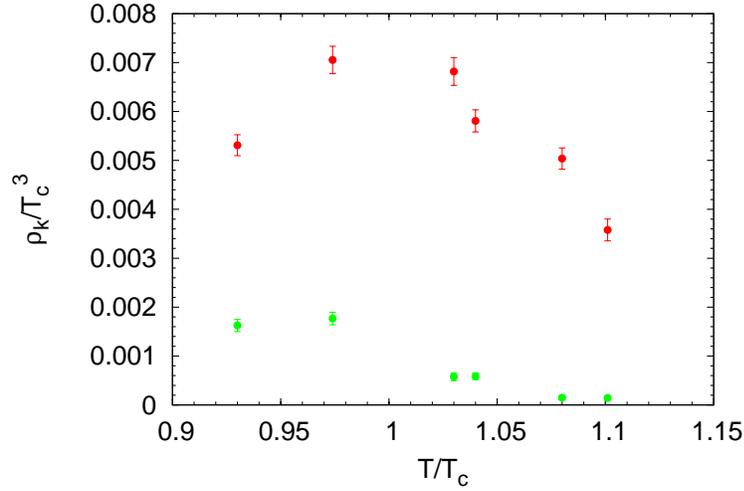}\\
\caption{Monopole density $\rho_k(T)$ for $k=2$ (red) 
and $k=3$ (green) in $SU(3)$
gluodynamics vs $T/T_c$. } \label{fig2}
\end{center}
\end{figure}

In Figure~\ref{fig1} we show density $\rho(T)$ defined in eq.~\ref{def_rho} 
and density $\rho_1(T)$ for thermal monopoles wrapped one time as function of
the ratio $T/T_c$ for temperatures below and above $T_c$. The data indicate 
volume independence of both densities. It can also be 
seen that $\rho_1(T)$ is a main contribution to $\rho(T)$ at all temperatures.
The important new observation is that both densities jump up at the transition
point. Thus density $\rho_1(T)$ is indicates the phase transition. One can
imagine that this sharp increase of $\rho_1(T)$ is due to evaporation
of the monopole condensate, existing in the confinement phase, or, in other 
words, many clusters with $N_{wr}=1$ appear from disintegrated percolating 
cluster. In Figure~\ref{fig1} we show for comparison the density $\rho(T)$
computed for full QCD. One can see that in this case the density also grows
fast near the phase transition.

The behavior of the densities $\rho_k(T)$ for $k=2$ and 3, shown in 
Figure~\ref{fig2}, is quite different. They increase below $T_c$ and decrease
above $T_c$.  

\begin{figure}[ht]
\begin{center}
\includegraphics[angle=270,scale=0.35]{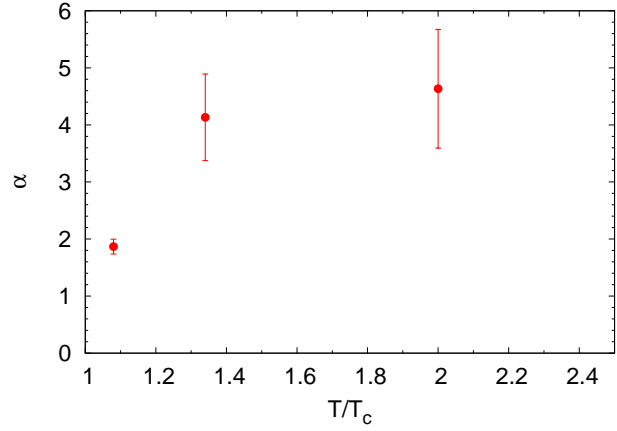}\\
\caption{Magnetic coupling $\alpha_M$ in SU(3) gluodynamics vs
$T/T_c$. } \label{fig3}
\end{center}
\end{figure}

\begin{figure}[ht]
\begin{center}
\includegraphics[angle=270,scale=0.35]{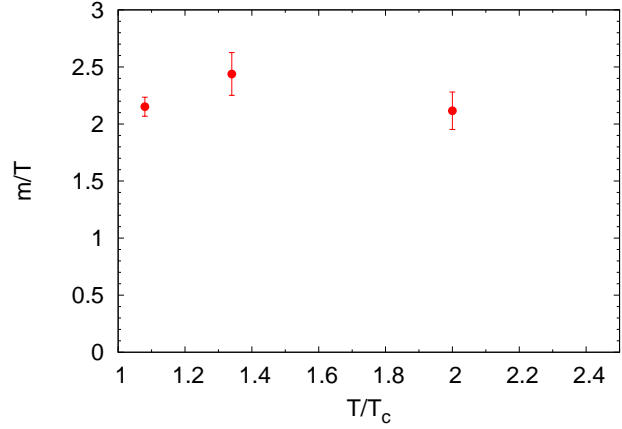}\\
\caption{Screening mass $m_D$ in SU(3) gluodynamics vs
$T/T_c$. } \label{fig4}
\end{center}
\end{figure}

We also computed the correlators for charges of same sign 
($g_{\tiny{MM}}(r)$) and  
for charges of opposite sign ($g_{\tiny{AM}}(r)$) :
\begin{equation}
g_{\tiny{MM}}(r) = \frac{\langle\rho^a_M(0)\rho^a_M(r)\rangle}{2\rho^b_M \rho^b_M} + \frac{\langle\rho^a_A(0)\rho^a_A(r)\rangle}{2\rho^b_A\rho^b_A}
\end{equation}
\begin{equation}
 g_{AM}(r) = \frac{\langle\rho^a_A(0)\rho^a_M(r)\rangle}{2\rho^b_A\rho^b_M} + \frac{\langle\rho^a_M(0)\rho^a_A(r)\rangle}{2\rho^b_A\rho^b_M}
\end{equation}
The correlators were fitted to functions \cite{D'Alessandro:2007su,Liao:2008jg}
\begin{equation}
 g_{MM,AM}(r)  = e^{-U(r)/T}\,,
\end{equation}
where
\begin{equation}
U(r)  = \frac{\alpha_{m}}{r}e^{-m_{D}r}
\end{equation}

In Figure~\ref{fig3} and Figure~\ref{fig4} we show dependence of the 
parameters $\alpha_{m}$ and $m_{D}$ on temperature. It can be seen that
$\alpha_{m}$ increases fast above $T_c$ and then flattens. $m_{D}/T$ increases
slightly near $T_c$ and starts to decrease at higher temperatures. These results are in agreement with results obtained for $SU(2)$ gluodynamics 
\cite{Liao:2008jg,Bornyakov:2011dj}.

\begin{figure}[ht]
\begin{center}
\includegraphics[angle=270,scale=0.35]{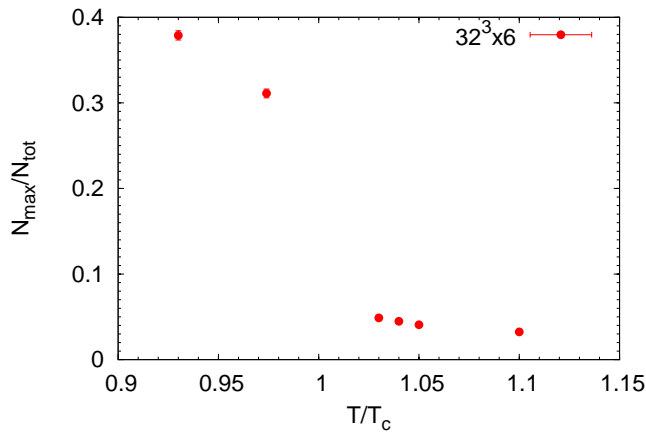}\\
\caption{Percolation transition order parameter. } \label{fig5}
\end{center}
\end{figure}

\begin{figure}[ht]
\begin{center}
\includegraphics[angle=270,scale=0.35]{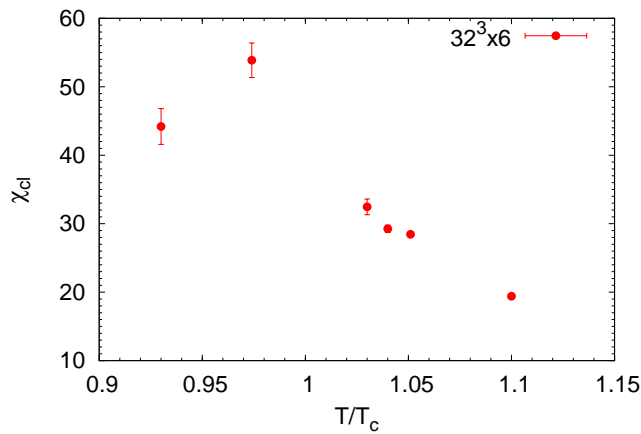}\\
\caption{Nonpercolating monopole cluster average size -
'susceptibility' $\chi_{cl}$}
\label{fig6}
\end{center}
\end{figure}

Additionally we have studied the percolation of the magnetic currents
at $T_c$ in $SU(3)$ gluodynamics. In Figure~\ref{fig5} the ration of the
average size of the maximal cluster to the full number of the magnetic 
currents, i.e. order parameter of the percolation transition 
is depicted. In  Figure~\ref{fig6} the nonpercolating cluster average size
which is often called susceptibility is shown. Both observables indicate
that the percolation transition coincides with the phase transition.

\section{Conclusions}
We have found that density of thermal monopoles both in 
$SU(3)$ gluodynamics and in QCD grows fast in the vicinity
of the transition (crossover) point. This is determined by
the $\rho_1(T)$ contribution alone.  Our data indicate 
volume independence of the densities $\rho(T)$ and $\rho_1(T)$. 
The magnetic coupling $\alpha_m$ and screening mass $m_D$ show
qualitatively same behavior as in $SU(2)$ gluodynamics. 

\section{Acknowledgements}
This work has been supported  by grant RFBR 13-02-01387-a.

\end{document}